\documentclass[sigconf]{acmart}
\usepackage{graphicx}
\usepackage{subcaption}
\usepackage{algorithm}
\usepackage{algorithmic}
\usepackage{siunitx}
\usepackage{float}
\usepackage{bbm}
\usepackage{balance} 
\usepackage{enumitem}
\usepackage{algorithm}

\AtBeginDocument{%
  }

\copyrightyear{2025}
\acmYear{2025}
\setcopyright{acmlicensed}
\acmConference[WWW Companion '25]{Companion
Proceedings of the ACM Web Conference 2025}{April 28-May 2, 2025}{Sydney, NSW, Australia}
\acmBooktitle{Companion Proceedings of the ACM Web Conference 2025 (WWW
Companion '25), April 28-May 2, 2025, Sydney, NSW, Australia}
\acmISBN{979-8-4007-1331-6/25/04}
\acmDOI{10.1145/3701716.3715256}
\begin{document}

\title[PRECTR: A Synergistic Framework for Integrating Personalized Search Relevance Matching \\ and CTR Prediction]{PRECTR: A Synergistic Framework for Integrating Personalized Search Relevance Matching and CTR Prediction}

\author{Rong Chen}
\affiliation{
  \institution{Alibaba Group}
  \city{Hangzhou}
  \country{China}}
\email{xingshu.cr@taobao.com}
\authornote{Both authors contributed equally to this research.}

\author{Shuzhi Cao}
\affiliation{
  \institution{Xi'an Jiaotong University}
  \city{Xi'an}
  \country{China}}
\email{cao309615@gmail.com}
\authornotemark[1]
\authornote{This work was done when Shuzhi Cao was a research intern at Alibaba Group. }

\author{Ailong He}
\affiliation{
  \institution{Alibaba Group}
  \city{Hangzhou}
  \country{China}}
\email{along.hal@taobao.com}

\author{Shuguang Han}
\affiliation{
  \institution{Alibaba Group}
  \city{Hangzhou}
  \country{China}}
\email{shuguang.sh@taobao.com}
\authornote{Corresponding Author}

\author{Jufeng Chen}
\affiliation{
  \institution{Alibaba Group}
  \city{Hangzhou}
  \country{China}}
\email{jufeng.cjf@taobao.com}

\renewcommand{\shortauthors}{Rong Chen, Shuzhi Cao, Ailong He, Shuguang Han, \& Jufeng Chen}

\begin{abstract}
The two primary tasks in the search recommendation system are search relevance matching and click-through rate (CTR) prediction -- the former focuses on seeking relevant items for user queries whereas the latter forecasts which item may better match user interest. 
Prior research typically develops two models to predict the CTR and search relevance separately, then ranking candidate items based on the fusion of the two outputs. 
However, such a divide-and-conquer paradigm creates the inconsistency between different models.
Meanwhile, the search relevance model mainly concentrates on the degree of objective text matching while neglecting personalized differences among different users, leading to restricted model performance.
To tackle these issues, we propose a unified 
\textbf{P}ersonalized Search \textbf{RE}levance Matching and \textbf{CTR} Prediction Fusion Model \textbf{(PRECTR)}. Specifically, based on the conditional probability fusion mechanism, PRECTR integrates the CTR prediction and search relevance matching into one framework to enhance the interaction and consistency of the two modules.
However, directly optimizing CTR binary classification loss may bring challenges to the fusion model's convergence and indefinitely promote the exposure of items with high CTR, regardless of their search relevance.
Hence, we further introduce two-stage training and semantic consistency regularization to accelerate the model's convergence and
restrain the recommendation of irrelevant items. 
Finally, acknowledging that different users may have varied relevance preferences, we assessed current users' relevance preferences by analyzing past users' preferences for similar queries and tailored incentives for different candidate items accordingly. Extensive experimental results on our production dataset and online A/B testing demonstrate the effectiveness and superiority of our proposed PRECTR method.
\end{abstract}



\maketitle

\section{Introduction}
Search relevance matching and click-through rate (CTR) prediction are the two core concepts in the search recommendation system. 
An ideal search recommendation system aims to present highly relevant items with elevated click-through rates to different users,  contingent upon their current queries and individual preferences.
Consequently, precisely modeling the relevance between queries and items and evaluating the CTR has emerged as the key challenge in the online search industry.

In the typical search recommendation system, two separate models are utilized to represent search relevance and CTR, respectively.
The search system will then make a comprehensive ranking of candidate items based on their outputs.
For example, Xianyu, the largest Chinese online second-hand trading platform, 
initially categorizes items into distinct relevance score levels based on their original relevance matching score and then ranks the candidate items according to their click-through rates within each level.
However, such distinct relevance score level separation results in somewhat less related commodities with high CTR losing exposure when they fall into the lower relevance score level.
Hence, the decoupling training of the two models resulted in the inconsistency between them, leading to restricted performance.
Meanwhile, current approaches merely depict relevance with the objective text matching degree and ignore the users’ personal relevance preferences towards various items.
All these factors constrain the performance of the current online search engine.
Therefore, it remains challenging to effectively integrate search relevance matching and CTR prediction into one overall framework to boost user interaction and personalize item sorting based on users' individual relevance preferences.

Recently, some pioneering works have attempted to capture the semantic relevance between the queries and items and take it as a supplementary feature for the CTR prediction model to generate a more comprehensive prediction.
For instance, \cite{SupplementaryNLPFeatures} employ a pre-trained language model \cite{bert,nlp1,nlp2} to extract the relevance feature from queries and items' raw texts and take it as an additional feature; 
\cite{numbert} transforms the basic numerical features into fine-grained tokens and feeds them with queries and items' textual features into the language model to boost their interaction.
On this basis, \cite{bert4ctr} further proposes Uti-Attention to avoid unnecessary interaction and 
reduce the computational complexity to meet the deployment requirements in real-world industrial scenarios.
Although all these methods successfully introduce the search relevance matching features into the CTR model to boost its performance, they only take it as a supplement for basic features and do not explicitly model the relevance between the queries and items, which restricts the capability of these models.

To resolve these issues,  we introduce a novel synergistic and unified Personalized Search Relevance Matching and CTR Prediction Fusion Model (PRECTR). 
Specifically, based on the conditional probability fusion mechanism, PRECTR integrates the CTR prediction and search relevance matching into one unified framework to enhance the feature interaction and module consistency.
However, directly optimizing the CTR binary classification loss poses challenges for the convergence of the fusion model and may lead to the overexposure of items with high CTR, irrespective of their search relevance.
Hence, we further introduce two-stage training and semantic consistency regularization to accelerate the model's convergence and restrain the recommendation of irrelevant items. 
Finally, recognizing that different users may have various levels of sensitivity to search relevance due to their personalized relevance preferences, 
we designed a personalized relevance incentive module to assess the current user’s relevance preference by analyzing her/his historical preference for similar queries and tailored incentives for different candidate items accordingly.

The main contributions of this paper are summarized as follows:

\vspace{-12pt}
\begin{itemize}
  \item To address the module inconsistency and insufficient interaction caused by decoupling training two separate models, we propose a novel synergistic Personalized Search Relevance Matching and CTR Prediction Fusion Model (PRECTR) to integrate CTR prediction and search relevance matching into one unified framework.
  \item The search relevance matching and CTR prediction are synergistically integrated under the conditional probability fusion mechanism. In addition, to tackle the problem of non-convergence for the fusion model and the overexposure of items with high CTR and extremely low search relevance, we introduce a two-stage training process and a semantic consistency regularization technique. Finally, realizing that users may have different tastes in semantic relevance, we analyze their historical preferences and offer tailored incentives accordingly. This integrated model alleviates the discrepancy between the two separate models and naturally improves the ranking performance.
  
  \item An extensive amount of experimental analysis on both industrial datasets and online A/B testing demonstrates the effectiveness and superiority of the proposed PRECTR method over the existing methods.
\end{itemize}

The remainder of this paper is organized as follows:
Section \ref{Related work} reviews related work.
Section \ref{PRELIMINARIES} describes preliminary concepts and notations. 
Section \ref{METHODOLOGY} details the proposed method.
Section \ref{EXPERIMENTS} presents experimental results.
Finally, Section \ref{conclusion}  concludes this paper.

\section{Related Work}
\label{Related work}
This section provides a concise overview of CTR prediction, search relevance matching, and their fusion techniques for final ranking.

\subsection{CTR Prediction}
CTR prediction plays a vital role in industrial search systems since it improves user experience and increases revenue for e-commerce platforms. Early statistics-based methods like collaborative filtering (CF) \cite{cf1,cf2,cf3} made recommendations by mining similarities between users or items. Subsequently, shallow machine-learning-based methods like logistic regression (LR) \cite{lr}, gradient boosting decision tree (GBDT) \cite{gbdt}, factorization machines (FM) \cite{fm}, field-aware factorization machines (FFM) \cite{ffm} are proposed to predict CTR.
The primary goal of these methods is to exploit feature interaction and examine how such interaction helps predict user clicks.

In recent years, the success of deep learning has brought the CTR prediction from shallow models to deep models, achieving remarkable performance.
Specifically, \cite{chen2016deep} first utilizes deep neural networks to predict CTR.
Wide\&Deep \cite{cheng2016wide} jointly trains wide linear models and deep neural networks to combine the benefits of memorization and generalization.
On this basis, DCN \cite{DCN} designs an additional cross-network to boost the feature interaction explicitly; DeepFM \cite{guo2017deepfm} combines the DNN and a factorization machine component to facilitate both high- and low-order feature interactions. Futhermore, AFM \cite{AFM} introduces the attention mechanism \cite{vaswani2017attention} to capture the importance of each feature interaction, leading to advanced CTR prediction performance.

Moreover, CTR models have become increasingly personalized.
To accommodate this trend of development, more and more powerful models \cite{DIN, DCNv2, DIEN, SIM,chang2023twin, lian2018xdeepfm, chen2024explicit, du2024disentangled, bst, Zhang2022TowardsUT, sheng2023joint, zhang2022keep, Gui2023Calibration} are proposed to capture the user's personal preferences by analyzing their historical behaviors.
For example, DIN \cite{DIN} uses Target-Attention to assess the relevance of the candidate item to previously clicked items for discovering the items that users are interested in.
On this basis, DIEN \cite{DIEN} further considers that user interest evolves dynamically and therefore designs an interest extractor layer to capture users' temporal interests. 
Moreover, BST \cite{bst} deploys the advanced Transformer \cite{vaswani2017attention} architecture into the CTR model to boost its prediction.
Later, the industry attempts to explore users' interests in long-sequence scenarios, and some representative work like SIM \cite{SIM} and TWIN \cite{chang2023twin} even extend the users' behavior sequence into lifelong scope, significantly improving the CTR prediction.

\subsection{Search Relevance Matching}
In e-commerce search systems, the relevance between the query and the item is usually measured by their text-matching score.
Therefore, search relevance matching is considered a text-matching task in most cases.
Early work such as TF-IDF similarity \cite{tf-idf1,tf-idf2} and Bag-of-Words (BoW) models \cite{bow1,bow2} perform keyword-based matching with statistical word frequencies.
Obviously, these methods fail to capture the contextual relationship among input texts thereby restricting the matching performance.
In recent years, more powerful methods built on top of deep learning techniques have been proposed for search relevance matching. They can be roughly divided into the interaction-based model \cite{hu2014convolutional,wang2022learning,qiao2019understanding,wan2016match, Han2020LearningtoRankWB} and the representation-based model.
The former puts all candidate texts together as inputs, then utilizes a pre-trained language model to extract their embeddings, and ultimately evaluates their relevance based on these representations.
Although these techniques show impressive performance, their high computational overhead makes them difficult for online deployment.
To tackle this issue, the representation-based model \cite{shen2014latent,guo2016deep,zhang2019improving,nigam2019semantic,yao2022reprbert,yao2021learning} is proposed to trade-off between performance and computation cost.
They encode the query and the item texts separately. This enables offline pre-computing with the sacrifice of text interaction. 

\subsection{The Fusion of CTR and Textual Relevance}
Ranking the candidate items by fusing both the estimated CTR and the textual relevance to the input query is a critical task for the search system. The current approach to integrating CTR and relevance can be divided into two groups.
The first group separately constructs two distinct models, one for CTR prediction and the other for computing relevance score. Then, the two scores are fused with human-defined strategies like hierarchical sorting or linear combination.
However, such a decoupled training process causes insufficient feature interaction, and inconsistency between the two models as well. To resolve this issue, later approaches \cite{bert4ctr, SupplementaryNLPFeatures, numbert} attempt to capture the relevance between queries and items and take it as a supplementary feature for CTR prediction. In this way, the model takes into account the two vital factors at the same time.
Nonetheless, a simple introduction of the relevance score as an input feature fails to explicitly model the relevance score, which may lessen the importance of such a feature.
Meanwhile, all these solutions overlook the fact that the impact of search relevance on a user's click probability is personalized.
Therefore, the fusion of CTR prediction and search relevance matching, and making further personalized recommendations based on user relevance preference remains a challenging problem.

\section{Preliminaries}
\label{PRELIMINARIES}
In this section, we first briefly formulate the problems of CTR prediction and search relevance matching and then introduce the existing CTR prediction model deployed in Xianyu.

\subsection{Problem Formulation}
Given the training dataset $\mathcal{D}=\left\{\boldsymbol{x},y\right\}^{n}$ collecting from our online e-commerce platform Xianyu, where $\boldsymbol{x}$ represents the high dimensional feature vector consisting of multi-fields features, the binary click label $y \in \left\{0,1\right\}$ indicates whether a sample is clicked, and $n$ denotes the number of training samples.
The feature $\boldsymbol{x}$ usually consists of multi-fields information and can be denoted as $\boldsymbol{x}=[t_{1}^\top,t_{2}^\top,...,t_{M}^\top]^\top$ where $t_{i} \in \mathbb{R}^{K_{i}}$ stands for the high-dimensional sparse binary features for $i$-th field, $K_{i}$ represents the dimensionality of vector $t_{i}$, and $M$ stands for the number of fields.
$t_{i}$ can be either a one-hot vector or a multi-hot vector depending on the number of values of the $i$-th field.
CTR prediction task aims to estimate the probability of a sample $x$ being clicked,i.e., $p_{ctr}(x)=p(y=1|x)$.

Similarly, the search relevance matching task can be formulated as below: given the original query text $Q_{text}$ and item description text $I_{text}$, our goal is to determine whether the two texts are relevant to each other.
In Xianyu, for each user query, all of the candidate items are divided into four different relevance score levels based on their relevance, i.e., the degree of textual matchness between the query and the item.

\subsection{CTR Prediction in Xianyu}
The Wide \& Deep model \cite{cheng2016wide, Wu2024MetaSplitMN,zhao2023entire} is currently employed by Xianyu as the base CTR prediction model. It jointly trains a wide linear model and a deep neural network which brings the benefits of memorization and generalization at the same time.
Here, we briefly introduce the underlying model architecture as follows:

\textbf{The Wide Component} is a generalized linear prediction model that can be expressed as $y=w^\top\boldsymbol{x}+b$. Here, $\boldsymbol{x}$ represents the input feature consisting of both raw features and the transformed cross-product features. $w$ and $b$ are model weights and biases respectively. The wide component takes advantage of the linear model and is thus incapable of memorizing the sparse feature interactions.

\textbf{The Deep Component} follows the Embedding and MLP paradigm, which first converts the sparse high-dimensional raw features into low-dimensional, dense real-valued vectors through the embedding layer and then utilizes MLP layers to extract their high-order representations. Such deep neural networks are easily generalizable to previously unseen feature interactions through low-dimensional embeddings, thus boosting the model's generalization ability. 
In this paper, we select the Wide\&Deep model as the base CTR prediction model for better aligning with our production setting.

\section{Methodology}
\label{METHODOLOGY}
This section outlines the overall design of our proposed method, including the architecture overview and the detailed structure.

\subsection{The Fusion Mechanism}
\label{Fusion Mechanism}
Intuitively, in a search system, a user clicking on an item is determined by its intrinsic quality and the relevance to the user query. Thus, as shown in the below Equation, we decompose the probability of clicking $P(click=1|\boldsymbol{x})$ on an item $\boldsymbol{x}$ into two parts: the relevance to the user query and the intrinsic item characteristic.
\begin{equation}
\label{eq1}
P(click=1|\boldsymbol{x})=\sum_{i=1}^{k}P(click=1|rsl=i,\boldsymbol{x})P(rsl=i|\boldsymbol{x}),
\end{equation}
Here, the variable $rsl$ stands for the relevance score level, which can be any of the $k$ discrete values. In specific, given a user query, all of the matching items in Xianyu are divided into four different relevance levels, i.e., $k\in\left\{1,2,3,4\right\}$, representing irrelevant, weakly-relevant, relevant, and strongly-relevant, respectively. 
$P(rsl=i|\boldsymbol{x})$ denotes the probability of the item $\boldsymbol{x}$ in the  relevance score level $i$ and $P(click=1|rsl=i,\boldsymbol{x})$ stands for the 
conditional probability of the item $\boldsymbol{x}$ being clicked in the $i$-th relevance score level.
To estimate the click probability in Eq.(\ref{eq1}), we design an additional Relevance Score Level Module (RSL Module for short) for predicting $P(rsl=i|\boldsymbol{x})$ and further combine it with the Base Module for predicting $P(click=1|rsl=i,\boldsymbol{x})$. Both of them output 4-dimensional vectors.
Subsequently, we take a dot product of their outputs according to Eq.(\ref{eq1}) to calculate the ultimate click probability $P(click=1|\boldsymbol{x})$ of the given item $\boldsymbol{x}$.

\subsection{The Model Input}
\label{Model input}
\textbf{The Base Module Input:} The Base Module takes all the features including the user, item,  context, and relevance features as the input to predict $P(click=1|rsl,\boldsymbol{x})$.

\noindent
\textbf{The RSL Module Input:} 
As for the RSL module, we first extract the relevance features $x_{rsl}$ that characterize the relevance relationship between the query words and candidate items as the input.
Specifically, $x_{rsl}$ is constructed as follows:
we use the original raw text of query words and item descriptions as the main input and construct derived features like "whether the category matches", "whether the description contains search terms" and so on to generate $x_{rsl}$.
Subsequently, we feed $x_{rsl}$ into the RSL module to estimate $P(rsl|\boldsymbol{x})$.
To handle the raw text feature, we utilize the pre-trained BERT to encode the text embedding.
Considering the latent high computational complexity resulting from high dimensional word representation, we also include a 3-layer MLP network for dimensionality reduction.
The pre-trained BERT and dimensionality reduction layers are trained offline in advance, including the following two stages:
\textbf{(1) Pre-trained Stage:} 
In this stage, we construct the training dataset via extensive available implicit feedback data collected from the Xianyu platform.
To be specific, we consider the clicked and unclicked query-item pairs as relevant and irrelevant text pairs, respectively, and pre-train the network through the binary classification task of identifying whether the text pair is related.
\textbf{(2) Fine-Tuning Stage:} In this stage, 
we manually annotate the relevance of the query-item pairs collected from the Xianyu platform and use it as supervised data to fine-tune the network.
However, directly integrating BERT into the CTR model may exacerbate the time-consuming challenge of online inference.
To solve this issue, we pre-compute the textual embeddings and store them in a static index table for online searching, maintaining low online inference latency.

\subsection{Two-Stage Training}
\label{Two-Stage Training}
Following the fusion mechanism in Section \ref{Fusion Mechanism}, we have successfully merged CTR prediction and search relevance matching within one unified training framework.
However, directly optimizing the CTR prediction binary classification loss through end-to-end training manner  
may bring challenges to the model's convergence.
This is because the click label is a too-weak supervised signal for all the parts of the model to converge to their true physical meaning, and the end-to-end loss function is usually more complicated, making the optimization process difficult.
Therefore, to realize the physical meaning of each module and accelerate the model's convergence,
we propose a two-stage training strategy.
Specifically, using the relevance score level $rsl$ as the supervised label, we first pre-train the RSL Module to initialize its parameters.
Suppose $T(\boldsymbol{x};\theta)$ represents the RSL Module parameterized by $\theta$, which models the probability distribution of different relevance score levels and can be expressed as $T(\boldsymbol{x};\theta)=[P(rsl=1|\boldsymbol{x}),..., P(rsl=4|\boldsymbol{x})]^\top$.
Taking the relevance score level $rsl$ as the supervised label, we warm up the parameters $\theta$ of the RSL Module by treating it as a multi-class classification task.
The corresponding pre-train risk is in cross-entropy form and is formulated as follows:
\begin{equation}
\label{eq2}
    \mathcal{R}_{pretrain}(\theta)=-\frac{1}{n}\sum_{i=1}^{n}rsl*log(softmax(T(\boldsymbol{x};\theta))),
\end{equation}
After the pre-training process, we then jointly train both the Base Module and the RSL Module by optimizing the CTR prediction objective.
It is worth noting that we apply a smaller learning rate for fine-tuning the RSL Module since the main goal is to update the parameters of the Base Module.
Let $g(\boldsymbol{x};\eta)$ denote the output of the Base Module, i.e, 
$g(\boldsymbol{x};\eta)=[P(click=1|rsl=1,\boldsymbol{x}),...,P(click=1|rsl=4,\boldsymbol{x})]^\top$.
The final click probability of the item $\boldsymbol{x}$ can be expressed as $P(click=1|\boldsymbol{x})=g(\boldsymbol{x};\eta) \cdot T(\boldsymbol{x};\theta)=f(x)$.
The co-training risk is defined as the pointwise CTR prediction risk in negative log-likelihood function, which is shown below:
\begin{equation}
\label{eq3}
    \mathcal{R}_{ctr}(\eta,\theta)=
    -\frac{1}{n}\sum_{(x,y)\in \mathcal{D}}(y\text{log}(f(x))+(1-y)\text{log}(1-f(x)),
\end{equation}

\subsection{Semantic Consistency Regularization}
\label{Semantic Consistency Regularization}
As mentioned in Section \ref{Two-Stage Training}, the fusion model first pre-trains the RSL Module and then takes the CTR prediction task as the final optimization objective for co-training. 
However, using the unitary click label for model training may inevitably lead to the recommendation of irrelevant cases.
The CTR prediction task tends to incentivize high click-through items to get exposure without considering their relevance to the user's query.
For example, when a user searches for "mobile phone", the search system may recommend unrelated "mobile phone cover" items as they have high click-through rates.

To restrain this phenomenon, we propose a novel listwise loss as the semantic consistency regularization to enhance the relevance standard of the recommended items.
Specifically, we design the listwise loss to keep the ranking order consistent with the ideal ranking order, where the ideal ranking of items is determined by both the item's click label and relevance score level.
The two main principles of ranking are shown as follows: (1) all the clicked items should be ranked preceding the unclicked items, and (2) the relevance priority should remain the same for the unclicked items, and thereby, the unclicked items with larger relevance score levels should be prioritized.

\begin{figure}
    \centering
    \includegraphics[width=0.98\hsize,height=0.335\hsize]{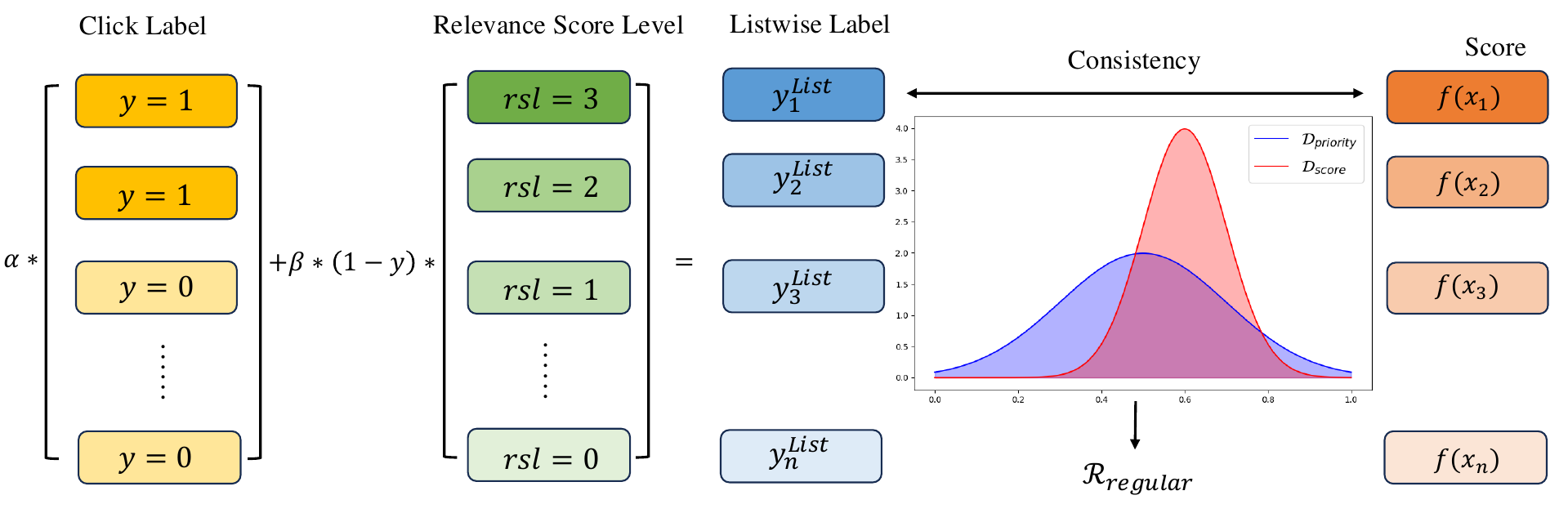}
    \caption{An illustration of the semantic consistency regularization.}
    \label{listwise}
    \vspace{-10pt}
\end{figure}

To achieve this goal, we re-defined the priority of items $\boldsymbol{x}$ by combining both their click label $y$ and their relevance score level $rsl$. To be specific, we synthesize a new listwise label $y^{List}$ to indicate item priority in a training batch, which is formulated as below:
\begin{equation}
\label{eq4}
    y^{List}=\alpha*y+\beta*(1-y)*rsl,
\end{equation}
Here $\alpha$ and $\beta$ are pre-defined hyper-parameters.
According to Eq.(\ref{eq4}), the larger the value of $y^{List}$, the higher the priority of the corresponding item $\boldsymbol{x}$.
We expect the distribution of the model scores for candidate items to be consistent with its distribution of priority.
Let $f(x)=P(click=1|\boldsymbol{x})=g(\boldsymbol{x};\eta) \cdot T(\boldsymbol{x};\theta)$ denote the final score for the item $\boldsymbol{x}$ given by the fusion model, $D_{score}=[f(x_{1}),...,f(x_{n})]$ and $D_{priority}=[y_{1}^{List},...,y_{n}^{List}]$ represent the score and priority distribution within a batch respectively.
We first use the softmax function to smooth these two distributions to 
ensure their comparability and subsequently minimize the KL divergence between them as an extra regularization to reinforce the semantic consistency, reducing the recommendation of irrelevant items.
The final semantic consistency regularization is formulated as follows:
\begin{equation}
\label{eq5}
    \mathcal{R}_{regular}=D_{KL}(softmax(D_{priority})||softmax(D_{score})),
\end{equation}
By adding the semantic consistency regularization in Eq. (\ref{eq5}), we introduce the relevance score level as an additional constraint to model training. This greatly reduces the recommendation of irrelevant items. The overview of semantic consistency regularization can be shown in Figure \ref{listwise}.

\subsection{Personalized Relevance Incentive}
In addition to the framework mentioned above that integrates CTR and textual relevance, we still omit that textual relevance can be personalized --- different users have different sensitivities to search relevance.
For example, users who are sensitive to textual relevance are less likely to click on irrelevant items, whereas the effect of query-item relevance may not affect that much on insensitive users.
Meanwhile, even for the same user, it may have different relevance tastes on different types of queries.
As a result, it is vital to model personal relevance preferences and provide individualized item recommendation experiences.
To achieve this goal, we include an extra \textbf{P}ersonalized \textbf{R}elevance \textbf{I}ncentive \textbf{M}odule \textbf{(PRIM)} in the training framework, which consists of three parts:

\textbf{Historical Relevance Preference Extraction Module:} We first extract the users' personal relevance preferences from their historical interaction records. 
Specifically, given a user's 
historical clicked item sequence and its corresponding query sequence 
$\mathcal{S}=\left\{(Q_{1}, I_{1}),...,(Q_{m}, I_{m})\right\}$, where $m$ denotes the length of historical click sequence, $Q_{i}$ and $I_{i}$ represent the raw text of the query and clicked item respectively. Then, we concatenate the original query and item text and add a special character [SEP] to mark their separation.
After that, we feed them as a whole sentence into the pre-trained language model and apply the MeanPooling operation in the output layer to aggregate the representation of the entire text as the relevance embedding $r^{emb}$.
Here $r^{emb}$ is regarded as the representation that contains information from both query and item text and is thus able to measure the degree of relevance.
This process is formulated as follows:
\begin{equation}
\label{eq6}
    r^{emb}_{i}=\text{Language\_Model}(\text{[CLS]}Q_{i}\text{[SEP]}I_{i}\text{[SEP]}).
\end{equation}

\textbf{Current Relevance Preference Estimation Module:}
In this module, we estimate the user's current relevance preference based on its immediate query words and relevance preference extracted from the historical click sequence.
As previously stated, different users may have varying relevance preferences towards various queries.
Therefore, to determine the user's relevance preference under the current query, we first assess the similarity between it and historical queries through the Target-Attention operation.
Let $Q_{cur}$ denote the raw text of the user's current query and $Q_{cur}^{emb}$ represent its embedding after the language model encoding.
Define $Q_{seq}^{emb}=[Q_{1}^{emb},..., Q_{m}^{emb}]$ and 
$r_{seq}^{emb}=[r_{1}^{emb},..., r_{m}^{emb}]$
the set of embedding vectors of queries and relevance preferences in the user's historical behaviors respectively, where $Q_{i}^{emb}$ denotes the embedding of raw query text $Q_{i}$. 
Taking current query $Q_{cur}^{emb}$ as query ($Q$), historical queries' embedding $Q_{seq}^{emb}$ as keys ($K$), and $r_{seq}^{emb}$ as values ($V$). 
Calculating Multi-Head Target-Attention (MHTA) operation based on given $Q$,$K$,$V$, the  user's current relevance preference expectation $r_{expect}^{emb}$ can be estimated as follows:
\begin{equation}
\label{eq7}
    r_{expect}^{emb}=\text{TargetAttention}(Q,K,V)=softmax(\frac{Q  K^\top}{\sqrt{d}})V,
\end{equation}
where $Q=w^{Q}Q_{cur}^{emb}$, $K=w^{K}Q_{seq}^{emb}$, $V=w^{V}r_{seq}^{emb}$, the matrices $w^{Q} \in \mathbb{R}^{d \times d}$, $w^{K} \in \mathbb{R}^{d \times d}$, and $w^{V} \in \mathbb{R}^{d \times d}$ are linear projection matrices, $d$ represents the dimension of feature embeddings.
The temperature $\sqrt{d}$ is introduced to produce a softer attention distribution for avoiding extremely small gradients.

\begin{figure*}
    \centering
    \includegraphics[width=0.96\hsize,height=0.48\hsize]{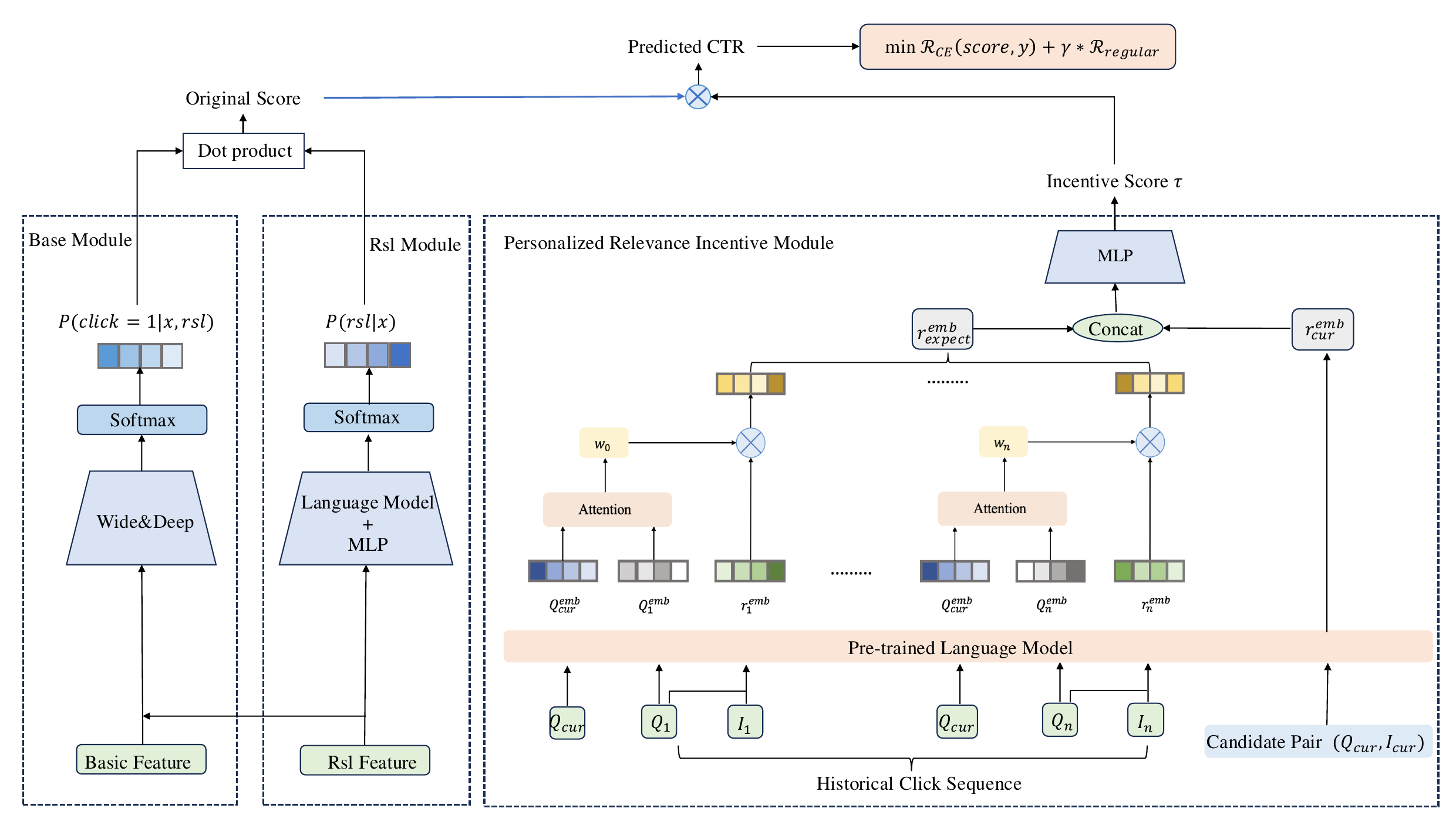}
    \caption{The overall framework of PRECTR can be broadly divided into three core components: the Base module, the Rsl module, and the Personalized Relevance Incentive module. The Base module and the Rsl module compute $P(click=1|x,rsl)$ and $P(rsl|x)$ respectively and end up with the original score, while the Personalized Relevance Incentive module computes the incentive score $\tau$ based on the user's personalized search relevance preference and finally generate the predicated CTR.}
    \label{overview}
\end{figure*}

\textbf{Personalized Relevance Incentive Module:}
After obtaining the user's current relevance preference estimation, 
we compute the relevance preference representation $r_{cur}^{emb}$ between the current query $Q_{cur}$ and the candidate item $I_{cur}$ with pre-trained language model.
Subsequently, we concatenate $r_{cur}^{emb}$ and $r_{expect}^{emb}$ together and feed it into a Multilayer Perceptron (MLP) network $M(;\omega)$ parameterized by $\omega$  to learn the incentive score.
Specifically, the MLP network outputs a scalar number $\tau$ that represents the intensity of the relevance preference, which is defined as follows:
\begin{equation}
\label{eq8}
    \tau=M(\text{concat}(r_{cur}^{emb},r_{expect}^{emb});\omega),
\end{equation}
Intuitively, $r_{cur}^{emb}$ stands for the relevance matching degree between the current query and the candidate item, and $r_{expect}^{emb}$ denotes the relevance expectation given the current query. 
As a result, the more similar the two representations are, the more the query-item candidate pairs align with the user's relevance preferences, and the greater the incentive score $\tau$ we should give.
On the contrary, the dissimilarity between the two representations indicates the inconsistency between the user's relevance preference and the current relevance standard.
In this case, the user may not click the candidate item due to its relevance preference and thus we should give a smaller incentive score $\tau$ to inhibit the recommendation of this item.
Finally, we use the incentive score $\tau$ to adjust the basic fusion score $g(\boldsymbol{x};\eta) \cdot T(\boldsymbol{x};\theta)$ and outputs the ultimate personalized score $score=\tau \cdot g(\boldsymbol{x};\eta) \cdot T(\boldsymbol{x};\theta)$.
Therefore, the final CTR optimization objective risk consists of two parts:
(1) a cross-entropy loss between the ultimate personalized score $score=\tau \cdot g(\boldsymbol{x};\eta) \cdot T(\boldsymbol{x};\theta)$ and click label $y$ and (2)
a semantic consistency regularization in subsection \ref{Semantic Consistency Regularization}, which is concluded as follows:
\begin{equation}
\label{eq9}
    \mathcal{R}(\eta,\theta,\omega)=\mathcal{R}_{CE}(score,y)+\gamma*\mathcal{R}_{regular},
\end{equation}
where $\mathcal{R}_{CE}$ denotes the cross-entropy risk and $\gamma$ is a pre-defined hyperparameter to trade-off between the cross entropy risk $\mathcal{R}_{CE}$ and regularization term $\mathcal{R}_{regular}$. 
For ease of understanding, we provide a visual representation of the proposed method's overview, shown in Figure \ref{overview}.

\section{Experiments}
\label{EXPERIMENTS}
In this section, we conduct comprehensive experiments on both the offline dataset and online A/B testing to evaluate the effectiveness and superiority of the proposed method.

\subsection{Experiments Setup}
\textbf{Dataset:} We collect click traffic logs from Alibaba's second-hand online trading platform Xianyu to construct the training dataset. 
Specifically, we treat clicked items as positive samples and unclicked items as negative samples. 
The daily training data volume is about 1.6 billion and each record contains 651 features  (e.g., user, query, and item features).
We take 9 days of data for the experiment evaluation,  in which data from the first 7 days is used for model training, and the remaining data of 2 days are used for model testing.

\textbf{Compared Methods:}  To demonstrate the superiority of the proposed method, we adopt the following baseline approaches:
\begin{itemize}
    \item \textbf{LR \cite{Logiticreg}:} Logistic Regression (short for LR) is a widely used shallow model before the deep model that estimates the CTR by combining features in a linear combination form.
    \item  \textbf{DNN:} DNN is proposed by YouTube and has been widely used in industry scenarios. It follows classic Embedding \& MLP architecture and utilizes the SumPooling operation to integrate historical behavior embeddings.
    \item \textbf{Wide\&Deep \cite{cheng2016wide}:} consists of both wide linear models and deep neural networks to benefit from memorization and generalization mutually. It is selected as our base model.
    \item \textbf{DeepFM \cite{guo2017deepfm}:} It combines factorization machine (FM) and deep neural network to improve Wide\&Deep, effectively capturing both low and high-order feature interactions.
    \item \textbf{XDeepFM \cite{lian2018xdeepfm}:} which proposes a novel Compressed Interaction Network (CIN) to generate feature interactions explicitly and combine it with DNN to predict CTR.
    \item \textbf{DIN \cite{DIN}:} which proposes a novel deep interest network and first utilizes the target attention operation to assess the relevance of the candidate item to previously clicked items for discovering the items that users are interested in.
    \item  \textbf{SuKD \cite{SupplementaryNLPFeatures}:} which utilizes the pre-trained BERT model to encode the raw text of query and item and take it as a supplementary feature to boost CTR prediction.
    
\end{itemize}
\begin{table*}[ht!]
    \centering
    \renewcommand{\arraystretch}{1.3}
    \setlength{\tabcolsep}{22pt}  
    \caption{The offline comparison results, where "RI" is short for "RelaImpr".}
     \begin{tabular}{c | c c |c c | c}
          \toprule[1.2pt]
          Method               &   AUC   & RI        &  GAUC    &   RI     & Relevance Score\\ \hline
          LR \cite{Logiticreg} & 0.6835  &  -27.36\% & 0.6230   & -33.31\% & 0.7188 \\
          DNN                  & 0.7528  & 0.06\%    & 0.6853	  & 0.47\%   & 0.7524 \\
          Wide\&Deep \cite{cheng2016wide} &0.7527	&0.00\%  & 0.6845	&0.00\%	 & 0.7538 \\
          DeepFM \cite{guo2017deepfm} &0.7526 &-0.04\% & 0.6852 &0.42\% & 0.7522\\
          XDeepFM \cite{lian2018xdeepfm} & 0.7517	&-0.39\%  &0.6839	&-0.29\%& 0.7501\\
          DIN \cite{DIN} &0.7546	&0.76\% &0.6870  &1.40\%  &0.7524\\
          SuKD \cite{SupplementaryNLPFeatures} &0.7513  &-0.54\% & 0.6841 &-0.22\%  & 0.7535\\
          \hline
          PRECTR   & $\boldsymbol{0.7548}$	& $\boldsymbol{0.82\%}$ & $\boldsymbol{0.6882}$ & $\boldsymbol{1.99\%}$ & $\boldsymbol{0.7561}$ \\
          \bottomrule[1.2pt]
    \end{tabular}
    \label{table1}
\end{table*}

\begin{table*}[ht!]
    \centering
    \renewcommand{\arraystretch}{1.3}
     \setlength{\tabcolsep}{19pt}  
    \caption{The ablation study of different PRECTR variants on production datasets, where "w/o" is short for "without".}
     \begin{tabular}{c c c c}
          \toprule[1.2pt]
          Method          &   AUC   & RI      &  Relevance Score\\ \hline
          Base            & 0.7538  & 0.00\%  &  0.7494              \\
          w/o Two-stage Training                   & 0.7628  & 3.53\%  & 0.7534\\
          w/o Semantic Consistency Regularization  & 0.7639  & 3.96\%  & 0.7559\\
          w/o Personalized Incentive               & 0.7640  & 4.01\%  & 0.7522\\
          \hline
          PRECTR  & $\boldsymbol{0.7642}$ & $\boldsymbol{4.04\%}$ & $\boldsymbol{0.7561}$ \\
          \bottomrule[1.2pt]
    \end{tabular}
    \label{table2}
\end{table*}

\textbf{Evaluation Metric:} The definition of the main evaluation metric used in the experiments is presented as follows:
\begin{itemize}[leftmargin=*]
  \item \textbf{AUC:} As the most commonly used evaluation metric in the search recommendation system, Area Under the Curve (AUC) reflects the sorting ability of the CTR model. 
  Specifically, given a positive and a negative item chosen randomly, AUC shows the likelihood that the model would rate the positive item higher than the negative one.
  Therefore, the AUC can be formulated as follows:
  \begin{equation}
      \text{AUC}=\frac{1}{|P||N|}\sum_{p\in P}\sum_{n\in N}\mathbbm{1}(\Theta(p)>\Theta(n)),
  \end{equation}
  where $P$ and $N$ represent the positive and negative item set respectively,  $\Theta$ is the ranking function given by the CTR model, and $\mathbbm{1}$ is the indicator function.
  \item \textbf{GAUC:} Different from the AUC that measures the global ranking ability of the model, the Group Area Under the Curve (GAUC) is designed to measure the goodness of order by ranking towards various groups or users. In specific, we first calculate AUC for different users and subsequently average them to get the final GAUC.
  The definition of GAUC is formulated as follows, where $\text{AUC}_{i}$ stands for the AUC for the $i$-th user, $\text{\#impression}_{i}$ is its corresponding weight, and $n$ denotes the total number of users.
  \begin{equation}
      \text{GAUC}=\frac{\sum_{i=1}^{n}(\#\text{impression}_{i}\times\text{AUC}_{i})}{
      \sum_{i=1}^{n}(\#\text{impression}_{i})
      }.
  \end{equation}
  \item \textbf{RelaImpr:} RelaImpr is adopted to measure the relative improvement over other models. $\text{RelaImpr}>0$ means the current model is superior over the base model, and vice versa. It can be formulated as follows:
  \begin{equation}
      \text{RelaImpr}=(\frac{\text{AUC}(\text{measured model})-0.5}{\text{AUC}(\text{base model})-0.5}-1)*100\%
  \end{equation}
  
  \item \textbf{Relevance Score:} To assess the relevance between the recommended items and the input query, for each query, we examine the relevance by intercepting the Top-10 items corresponding to the recommendations. 
  To be specific, we utilize the pre-trained language model mentioned in subsection \ref{Fusion Mechanism} to extract query and item's embedding $Q^{emb}$ and $I^{emb} $ respectively.
  Afterward, we calculate the cosine similarity between them to reflect their similarity. 
  By averaging the similarities of all query-item pairs, we get the model's relevance score below, where $L$ denotes the number of queries in total.
  \begin{equation}
      \text{Relevance Score}=\frac{1}{10L}\sum_{i=1}^{L}\sum_{j=1}^{10}
      \frac{Q_{i}^{emb} \cdot I_{j}^{emb}}{||Q_{i}^{emb}|| \times ||I_{j}^{emb}||}
  \end{equation}
\end{itemize}

\textbf{Implementation Details:} In all of the experiments, the batch size is set to $4096$, and the SGD optimizer is employed for model update.
In the pre-train stage, we first warm up the RSL Module under the supervision of relevance score level with an initial learning rate 1e-4.
As for the co-training stage, we fine-tune the RSL Module by adjusting the learning rate to 1e-5 while the other modules remain the same.
The history click sequence is collected with the last 30 days and the maximum length is 50.


\subsection{Experimental Results}
In this subsection, we discuss the experimental results from the
following three aspects:

\textbf{(1) Offline Experimental Results:} To demonstrate the effectiveness and superiority of the proposed method, we conduct offline experiments on the Xianyu dataset and compare it with state-of-the-art (SOTA) approaches. 
For a fair comparison, we cold start all of the models and train them from scratch.
The results are presented in Table \ref{table1}. 
Compared with the previous approaches, our method achieved the best performance on all of the metrics.
Specifically, compared to the base model Wide\&Deep, our approach achieved a $0.82\%$ improvement in AUC and a $1.99\%$ increase in GAUC.
Moreover, our model is also capable of perceiving the search-matching relevance and achieves the best $0.7561$ relevance score that greatly exceeds the compared methods.

\textbf{(2) Ablation Experiments:} We conduct ablation experiments to verify the effectiveness of all the modules and strategies proposed in our method.
Take the Wide\&Deep model as the base model which directly estimates the CTR without considering the search matching relevance.
To verify each module's contribution, we compare the complete PRECTR method with the model without two-stage training, the model without semantic consistency regularization, and the model without personalized relevance incentive module, respectively.
Unlike offline experiments, we use the online model to warm up our model and inherit the knowledge learned in history to simulate the online environment.
The corresponding results are summarized in Table \ref{table2}.
Specifically, compared with the base model, the complete PRECTR model achieves both the best AUC of $0.7642$ and the best relevance score of $0.7561$, while the base model shows the worst performance.
Meanwhile, various ablation versions of models gain in AUC and relevance scores compared with the base model, exhibiting the effectiveness of each module and strategy proposed in our method.
Nonetheless, we also observe that the semantic consistency regularization does not greatly enhance the overall model's performance, this might be because the majority of training data are strongly relevant items, weakening the influence of the semantic consistency regularization.

\textbf{(3) Online A/B Testing:} We deploy the proposed model online in Xianyu's search recommendation system to test its online performance. 
Through an online A/B test, users are randomly assigned to control and experimental groups. 
Device IDs are uniformly distributed via MD5 hashing for impartial partitioning. 
We observed the online results in the experimental group for 7 days, which had over $5\%$ of the total traffic.
Specifically, the proposed PRECTR model improved across key
metrics compared to the base model, with a $0.4\%$ increase in CTR
and a $1.1\%$ increase in Gross Merchandise Volume (GMV).
Meanwhile, the consistency between the CTR and CVR is further improved, the CTCVR metric achieves a $1.04\%$ improvement compared to the current online serving model.


\subsection{Hyperparameter Tuning}
In this subsection, we conduct experiments to select the optimal hyperparameters for the proposed method.

\begin{figure}
    \centering
    \includegraphics[width=0.92\linewidth]{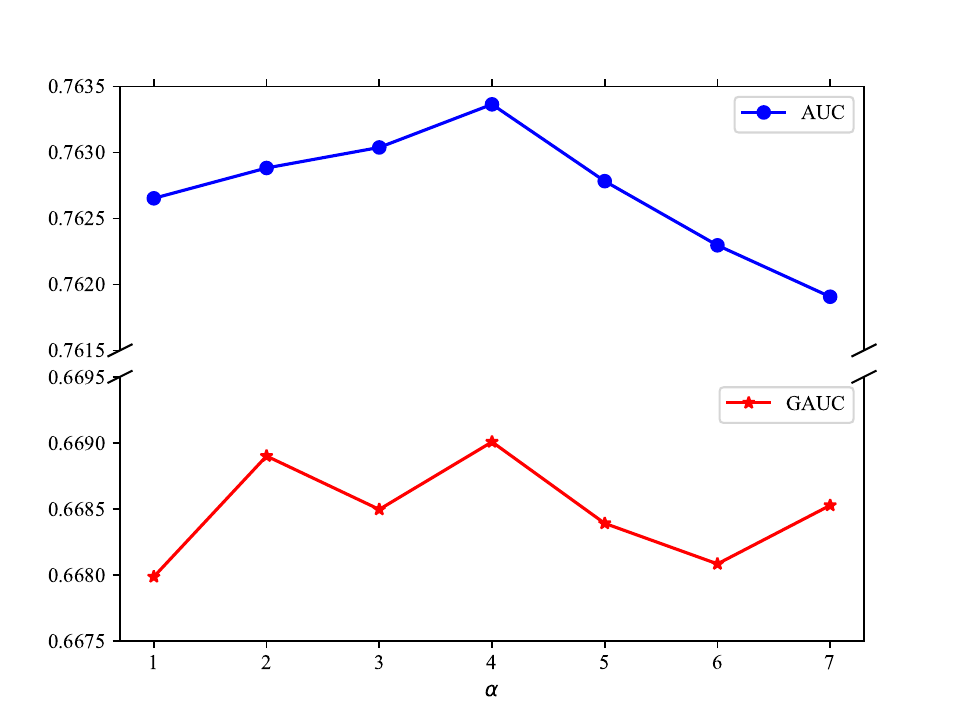}
    \caption{The effect of the hyperparameter $\alpha$ in semantic consistency regularization.}
    \label{ablation}
    \vspace{-15pt}
\end{figure}

\textbf{(1) The selection of $\alpha$ and $\beta$ in semantic consistency regularization:}
$\alpha$ and $\beta$ represent the weight of click label $y$ and relevance score level $rsl$ respectively.
To effectively search for their optimal value, we set $\beta=1$ as a constant and change different values for $\alpha$ to observe its effect on the ultimate AUC and GAUC.
According to Figure \ref{ablation}, with the rise of $\alpha$, the AUC and GAUC roughly rise at first and then drop.
When $\alpha=4$, our method achieves the best AUC and GAUC performance at the same time.
Therefore, we select $\alpha=4$ as the optimal hyperparameter value in all the experiments.

\begin{figure}
    \centering
    \includegraphics[width=0.92\linewidth]{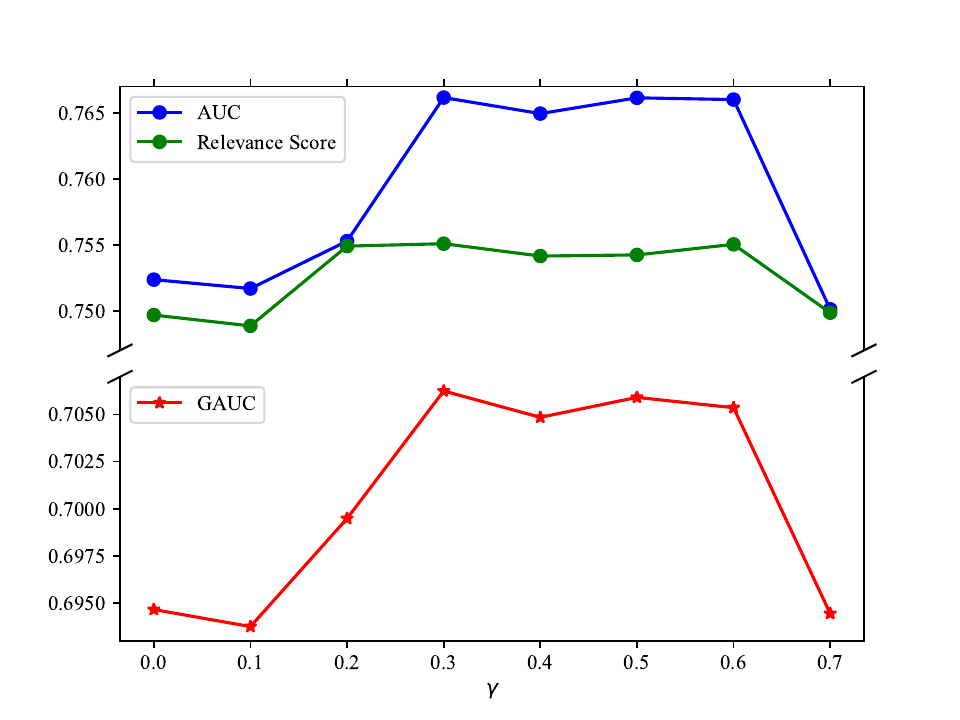}
    \caption{The effect of the hyperparameter $\gamma$ in the final optimization objective.}
    \label{ablation2}
    \vspace{-15pt}
\end{figure}

\textbf{(2) The selection of $\gamma$ in the final optimization objective:}
According to Eq.(\ref{eq9}), the hyperparameter $\gamma$ trades off between the CTR binary classification risk and semantic consistency regularization risk. 
To determine its optimal value, we iterate various values from small to large to test the model's performance.
As shown in Figure \ref{ablation2}. , when $\gamma$ is set to $0.3$, the overall model achieves the highest AUC, GAUC, and Relevance Score at the same time. 
Therefore, we set $\gamma=0.3$ in all the experiments.

\section{Conclusion}
\label{conclusion}
In this paper, we propose a synergistic framework called PRECTR to address inconsistencies from decoupling the training of separate models for search relevance and CTR prediction.
Based on the conditional probability fusion mechanism, PRECTR integrates the CTR prediction and relevance modeling into one unified framework.
Furthermore, we introduce the two-stage training and semantic consistency regularization to boost their integration.
Finally, by analyzing the users' historical click sequences, we design an additional 
personalized relevance incentive module to offer tailored incentives for different users.
We conduct comprehensive experiments on the Xianyu datasets and online A/B testing to demonstrate the effectiveness and superiority of our proposed method.

\bibliographystyle{ACM-Reference-Format}
\balance
\bibliography{sample-base}

\end{document}